\documentclass[twocolumn,showpacs,preprintnumbers,amsmath,amssymb]{revtex4}

\usepackage{graphicx}
\usepackage{dcolumn}
\usepackage{bm}
\usepackage{subfigure}

\begin{document}

\title{Electron and Photon Interactions in the Regime of Strong LPM Suppression}

\author{Lisa Gerhardt} \affiliation{Lawrence Berkeley National Laboratory, Berkeley CA, 94720 
USA \\ and Department of Physics, University of California, Berkeley, CA, 94720 USA}
\author{Spencer R. Klein} \affiliation{Lawrence Berkeley National Laboratory, Berkeley CA, 94720 
USA \\ and Department of Physics, University of California, Berkeley, CA, 94720 USA}

\date{\today}

\begin{abstract}

Most searches for ultra-high energy (UHE) astrophysical neutrinos look for radio emission from the electromagnetic and hadronic showers produced in their interactions.  The radio frequency spectrum and angular distribution depend on the shower development, so are sensitive to the interaction cross sections.     At energies above about  $10^{16}$ eV (in ice), the Landau-Pomeranchuk-Migdal (LPM) effect significantly reduces the cross sections for the two dominant electromagnetic interactions: bremsstrahlung and pair production.   At higher energies, above about $10^{20}$ eV,  the photonuclear cross section becomes larger than that for pair production, and direct pair production and electronuclear interactions become dominant over bremsstrahlung.  The electron interaction length reaches a maximum around $10^{21}$ eV, and then decreases slowly as the electron energy increases further.   In this regime, the growth in the photon cross section and electron energy loss moderates the rise in $\nu_e$ shower length, which rises from $\sim$10 m at $10^{16}$ eV to $\sim$50 m at $10^{19}$ eV and $\sim$100 m at $10^{20}$ eV, but only to $\sim$1 km at $10^{24}$ eV.  In contrast, without photonuclear and electronuclear interactions, the shower length would be over 10 km at $10^{24}$ eV.

\end{abstract}

\maketitle

\section{Introduction}

Many groups are searching  for astrophysical neutrinos with energies above 100 PeV.   Experiments that search
for optical Cherenkov radiation have sub-TeV thresholds \cite{IceCube} and detectors with radio antennas emplaced in
Antarctic ice have thresholds around $10^{17}$ eV \cite{RICE,ARIANNA, ARA}, but most of the other experiments are only sensitive to considerably higher energies.  For example, ANITA \cite{ANITA} is most sensitive above $10^{19}$ eV, while experiments that search for acoustic signals \cite{Acoustic} or for neutrino interactions in the Moon have threshold above $10^{20}$ eV \cite{Parkes,GLUE,numoon,Lunaska,RESUN,Lofar} - often far above.  These experiments observe the 
coherent radio Cherenkov radiation or acoustic emission from the showers produced by $\nu_e$ and neutral current interactions of all neutrino flavors, and have presented flux limits on the cosmogenic neutrino flux at energies up to $10^{25}$ eV. 

Neutrino induced showers have hadronic and electromagnetic components.  The hadronic component comes from the average of 20\% of the neutrino energy that is transferred to the target nucleus. The LPM effect has little effect on hadronic showers, so these showers remains relatively compact, less than 40 m long in ice.   

In charged-current interactions the remainder of the energy is transferred to the produced lepton.  For $\nu_e$ ($1/3$ of the total neutrino flux), this energy produces an electromagnetic shower, which also produces radio and acoustic radiation.  Because of the higher shower energy, this electromagnetic component is potentially easier to detect than the hadronic component.   However, the LPM effect greatly elongates the shower; for a $10^{24}$ eV neutrino interacting in ice,  the average shower length is over 10 km \cite{klein04}.   This length restricts the conditions for coherent radiation to a very narrow angular range, limiting its usefulness. 

Further, because soft bremsstrahlung is suppressed, the energy deposition from the shower is `clumpy', with large shower-to-shower variations \cite{konishi91}, leading to large event-by-event differences in radio emission.   These effects are difficult to implement and most of the experiments with high thresholds have ignored electron initiated showers entirely, calculating their sensitivity to $\nu_e$ interactions solely from the hadronic showers.   

In this paper, we present more complete calculations of electron and photon interaction cross sections, including new effects that suppress the bremsstrahlung cross section at high energies. We consider photonuclear and electronuclear cross sections, along with direct pair production; these processes dominate over pair production and bremsstrahlung at energies above $10^{20}$ eV.   Exact calculations of these processes do not exist, so we present computations of a phenomenological nature. We determine the total electron and photon interaction lengths and use a simple model to calculate the shower length as a function of neutrino energy. Above the energy where photonuclear and electronuclear interactions dominate,  the shower length only rises modestly with further increases in neutrino energy \cite{ralston02}. 

\section{Cross Sections with LPM and Dielectric Suppression}\label{sect:two}

The bremsstrahlung and pair production cross sections were first calculated by Bethe and Heitler \cite{bet43}.  A  modern, detailed rendition of the Bethe-Heitler cross section ($\sigma_{BH}$) for an electron with energy E to radiate a photon with energy $k= yE$ is \cite{tsai74}
\begin{eqnarray}
\frac{d\sigma_{BH}} {dk} = 
\frac{4\alpha r_e^2}{3k}
\big(   
(y^2+2[1+(1-y)^2])   \quad \quad \quad \\
\times \big(Z^2\ln{(184 Z^{-1/3})}\!+\! Z\ln{(1194Z^{-2/3})}\big)
 +  (\!1\!-y) \frac{Z^2+Z}{3}
\big) \quad \;
\label{eq:BH}
\end{eqnarray}
This is in the ``full-screening'' limit, appropriate at high energies.  Only the first logarithmic term, for electrons interacting with the target nucleus, was considered by Bethe and Heitler.  The second logarithmic term is for electron interactions with atomic target electrons.  This equation is simplified by introducing the radiation length, $X_0$ and neglecting the different $y$ dependence of the last term:
\begin{equation}
{d\sigma_{BH}\over dk} = {1\over 3nkX_0}
(y^2+2[1+(1-y)^2])
\label{eq:BH2}
\end{equation}
where $n$ is the number of targets per unit volume.  Similarly, the cross section for a photon with energy $k$ to produce an $e^+e^-$ pair with lepton energies $E$ and $k-E$ is 
\begin{equation}
{d\sigma_{BH}\over dE} = {1\over 3nkX_0}
(1+2[x^2+(1-x)^2])
\label{eq:BH3}
\end{equation}
where $x=E/k$.

Landau and Pomerachuk \cite{land53} showed that, at high energies, the bremsstrahlung and pair production cross sections are suppressed.  The longitudinal momentum transfer between the bremsstrahlung electron and the target nucleus is small, $q_{||} = km^2c^3/2E(E-k)$ for an electron with mass $m$ energy $E$ to emit a photon with energy $k$.   For pair production, $q_{||}= km^2c^3/2E(k-E)$, where $k$ is again the photon energy, and $E$ the electron energy.    

The interaction is delocalized over a formation length, $l_f=\hbar/q_{||}$, and one adds the amplitudes for all of the electron-target interactions within that distance. When the incident particle energy is large enough so that $l_f$ is larger than the inter-atomic spacing, destructive interference can reduce the cross section.  Semi-classically, if there are enough interactions in this distance to scatter the electron significantly (by an angle larger than $m/E$), then radiation is
reduced \cite{anthony9598, hansen04}.   For bremsstrahlung, this suppression is significant when
\begin{equation}
k < k_{LPM} =\frac{E(E-k)}{E_{LPM}}.
\label{eq:LPMonset}
\end{equation}
$E_{LPM}$ is a material dependent constant
\begin{equation}
E_{LPM} = {m^2c^3 X_0 \alpha \over 4\pi\hbar} =  7.7\  {\rm \frac{TeV}{cm}} \cdot X_0.
\end{equation}
For ice (with a density 90\% that of water), $E_{LPM} =   303$ TeV, while in standard rock, with a density of 2.65 g/cm$^3$ and $X_0=10$ cm, $E_{LPM}= 77$ TeV.    The degree of suppression may be calculated semi-classically:  one singles out a single target as causing the bremsstrahlung, with the other targets contributing via multiple scattering.  This approach is in reasonable agreement with more accurate  calculations.   For $k<E(E-k)/E_{LPM}$, the suppression is found by calculating $q_{||}$, including a term to account for the change in electron momentum due to the multiple scattering; the change in direction reduces its forward momentum \cite{spen99}, and 
\begin{equation}
q_{||} = {km^2c^3\over 2E(E-k)} + {k \theta_{MS2}^2\over 2c}.
\label{eq:qperpdiel}
\end{equation}
where $\theta_{MS2}$ is the angle that the electron scatters in traversing half of the formation length $l_f/2$.  
For multiple Coulomb scattering, the mean scattering angle for a relativistic electron in a distance $l_f/2$ is
\begin{equation}
\theta_{MS2} ={ E_s \over E} \sqrt{{l_f\over 2X_0}}
\end{equation}
where $E_s = mc^2\sqrt{4\pi/\alpha} = 21.2$ MeV; $\alpha\approx 1/137$ is the electromagnetic coupling constant.   $\theta_{MS2}$ depends on $l_f$, and $q_{||}=\hbar/l_f$ depends on $\theta_{MS2}^2$, so one obtains a quadratic equation for $l_f$.  

The degree of suppression, $S$, is the ratio of the reduced formation length to the formation length in open space:
\begin{equation}
S = {d\sigma_{LPM}/dk \over d\sigma_{BH}/dk } = \sqrt{kE_{LPM} \over E(E-k)}
\label{eq:LPsuppression}
\end{equation}
Low energy radiation is reduced the most.

The semi-classical calculation is easily expanded to include new phenomena by adding additional terms to $q_{||}$.  For example, Ter-Mikaelian described the longitudinal density effect (also known as the dielectric effect), which reduces bremsstrahlung when $y<\approx 10^{-4}$ \cite{ter53}.  It arises from interactions between the emitted photon and the target medium.  The photon acquires an effective mass of $\hbar\omega_p$, where $\omega_p$ is the plasma frequency of the medium.  $\omega_p$ depends on the target electron density.  In typical solids $\hbar\omega_p$ is of order 10-60 eV.  Dielectric suppression can be calculated semi-classically by adding a term to $q_{||}$:
\begin{equation}
q_{Di}=\frac{(\hbar\omega_p)^2}{2ck}
\end{equation}
In the absence of other suppression mechanisms, bremsstrahlung is suppressed by $(k/\gamma\hbar\omega_p)^2$ when $k< \gamma \hbar\omega_p$.   When the LPM effect is included, dielectric suppression dominates when \cite{feinberg56}
\begin{equation}\label{eqtn:kdie}
k\le\sqrt[3]{\frac{(E\hbar\omega_p/mc^2)^4} {k_{LPM}}} 
\end{equation}
For ice with $\hbar\omega_p\approx20$ eV, this is
\begin{equation}
y  < 1.3\times10^{-6} \times \sqrt[3]{\frac{E_{LPM}}{E}};
\end{equation}
as $E$ rises and LPM suppression becomes more important, this mechanism loses importance. 

For pair production, the suppression for $k>E_{LPM}$ is
\begin{equation}
S = {d\sigma_{LPM}/dE \over d\sigma_{BH}/dE } = \sqrt{kE_{LPM} \over E(k-E)}
\end{equation}
For equal energy electrons and photons, the photon cross sections are much less suppressed than the electron $d\sigma/dk$.  The overall (integrated over $E$) pair production cross section is reduced by 50\% for $k\approx 70 E_{LPM}$. In contrast, electron $dE/dx$ is reduced by 50\% when $E=E_{LPM}$; in a shower, the LPM effect is much larger for electrons than photons. 

In  1956, Migdal gave a fully quantum mechanical calculation of the cross sections at high energies, including multiple scattering and the dielectric effect \cite{mig56}.  For bremsstrahlung,
\begin{eqnarray}
{d\sigma_{LPM} \over dk}\! &\! =\!& {4{\alpha}r_{e}^2\xi(s) \over 3k} 
\big(y^2G(s)\! +\!2\![\!1\!+\!(1\!-\!y)^2\!]\!{\phi(s\Gamma)\over \Gamma} \big) \\
&& \times Z^2 \ln{(184Z^{-\!1\!/\!3\!})}
\label{eqn:migdie}
\end{eqnarray}
where $\Gamma= 1 + (\gamma\hbar\omega_{p}/k)^2$.

Migdal neglected the electron-electron scattering, but this can be included (assuming that it is suppressed in the same manner as electron-nucleus interactions) by renormalizing this into the radiation length. For bremsstrahlung
\begin{equation}
{d\sigma_{LPM} \over dk}  ={ \xi(s) \over 3nX_0k} 
\big(y^2 G(s) + 2[1+(1-y)^2] {\phi(s\Gamma)\over\Gamma} \big)
\label{eq:LPM}
\end{equation}
while for pair production:
\begin{equation}
{d\sigma_{LPM} \over dE}  ={ \xi(s) \over 3nX_0k} 
\big(G(s) + 2[x^2+(1-x)^2]\phi(s)\big)
\label{eqn:migpair}
\end{equation}

Here, $s$ is related to the degree of semiclassical suppression, defined as Eq. (\ref{eq:LPsuppression}) divided by $\sqrt{8\xi(s)}$.  The functions $G(s)$ and 
$\phi(s)$ give the degree of suppression.   The factor $\xi(s)$ varies logarithmically as $s$ rises within the limits $1 \le \xi(s) \le 2$.

Migdal provided recursive formulae for $G(s)$ and $\phi(s)$.  Greatly simplified, non-recursive formulae were developed by Stanev and collaborators \cite{stan82}. For all of these equations, the Bethe-Heiter equivalent is obtained by setting $G(s)=\phi(s)=\xi(s)=1$. 

The Migdal cross sections are used in most calculations of shower production in neutrino interactions.  Although more sophisticated calculations exist \cite{baier00,zakharov99}, their predictions are similar to Migdal's.  In the strong suppression limit, Zakharov's calculations \cite{zakharov99} should lead to a slightly lower cross section than Migdal \cite{klein06}, while Baier and Katkov \cite{baier03} find pair production cross sections that are 2-10\% percent larger than Migdal \cite{klein06}.

\section{Further Bremsstrahlung Cross Section Suppression}
\label{sect:three}
Another suppression mechanism arises when the bremsstrahlung formation length becomes larger than the interaction length of the created photon, $l_I$.  A newly created bremsstrahlung photon may interact before it forms fully; the photon amplitude from the beginning of the formation length cannot be added to that at the end, because the photon amplitude from the beginning decays away (due to its interaction probability) before the end \cite{gal64}.  The photon can interact via either pair production or a photonuclear reaction.  This suppression can be included in the semi-classical calculation by invoking the uncertainty principle and adding a term to $q_{||}$:
\begin{equation}
q_{I}=\frac{\hbar}{l_I} = {\hbar}n(\sigma_{PP} + \sigma_{\gamma A})
\end{equation}
where $\sigma_{PP}$ is Migdal's pair production cross section  [Eq. (\ref{eqn:migpair})] and $\sigma_{\gamma A}$ is the photonuclear interaction cross section (described in Section \ref{sect:photon}). 

For photons with energies less than a few $E_{LPM}$, the interaction length is constant, $(9/7) X_0$, and this mechanism becomes visible (on top of LPM and dielectric suppression) when \cite{gal64}
\begin{equation}
E >  {2X_0\omega_p E_s\over c}
\end{equation}
Other analyses have found slightly different coefficients \cite{spen99}. 
For ice (standard rock), this is $E>1740$ TeV (645 TeV), beyond the reach of accelerator experiments, but very relevant for UHE astrophysical neutrinos.  Photonuclear interactions can similarly limit the bremsstrahlung formation length, reducing radiation by electrons \cite{ralston02}.

At higher photon energies, the LPM effect increases the photon pair production length $l_{PP}=\hbar/\sigma_{PP}$.  When $l_{PP}$ becomes longer than the electron interaction length,  bremsstrahlung from the partially produced electron and positron can limit $l_{PP}$.    When $l_{PP}$ is longer than the electron interaction length, the bremsstrahlung probability depends on the pair production probability, which in turn depends on the bremsstrahlung probability.  Electronuclear and photonuclear interactions, discussed below, can also limit the formation length, further complicating the problem.  Since the particle energy decreases as one moves down the chain of interactions,  the cross sections could be found iteratively, starting with low energy bremsstrahlung and pair production.   

We use a simpler approach, taking advantage of the fact that, at a given energy,  the LPM effect is much less important for photons, and  that photonuclear interactions become important when the pair production cross section is strongly suppressed.  We start with a calculation of photon interaction probabilities, and then use that to find the bremsstrahlung cross sections, non-iteratively.  In doing so, we neglect the effect that bremsstrahlung from one of the produced leptons has on pair production; this should be small because of the longer electron interaction lengths. 

As the photon and electron energies rise, the cross sections drop and the interaction length increase as the square root of the particle energy.  When $S$ is small enough ($\approx \alpha$), it is natural to expect that higher order processes  like electronuclear interactions and direct pair production $e^-N\rightarrow e^+e^-e^-N$  to become important.  

\section{Photon interaction cross sections}\label{sect:photon}

At low energies, photons interact via pair production.  At higher energies, pair production is LPM-suppressed and photonuclear interactions become important. At an energy above about $10^{20}$ eV (surprisingly independent of the target material), photonuclear cross sections are larger than for pair production \cite{klein04,klein06}.  There is significant uncertainty in extrapolating measured  photonuclear cross sections to high energies; we use the Engel parameterization \cite{engel97,couderc09}. It uses generalized vector dominance plus a term for direct photon-quark interactions \cite{engel97}.  In the high-energy regime, the cross section rises as $k^{0.08}$.  For nuclei heavier than protons, the cross section scales with atomic number $A$ as $A^{0.887}$.  These cross sections are similar to those found in a dipole model \cite{rogers06}. On the other hand, calculations based on a Glauber model \cite{KN99} give cross sections that are $\sim$2 times lower \cite{couderc09}.

At still higher energies photons interact coherently with the target medium as a whole.  The photons fluctuate to quark-antiquark pairs which elastically scatter from target nuclei, emerging as real $\rho^0$ mesons. At energies above $10^{23}$ eV, the photon to $\rho^0$ conversion has a long formation length, so the process is delocalized and the amplitudes add coherently \cite{couderc09}.   The $\rho^0$ will interact hadronically, or decay to $\pi^+\pi^-$, which will themselves interact hadronically.   This photon to $\rho^0$ conversion is important at the upper end of the energy range considered here, where it will further reduce the photon interaction length. However, because of the uncertainties in the cross-section, we do not include it in our numerical estimates.  

\begin{figure} [tb]
\centering
\includegraphics[totalheight=0.3\textheight]{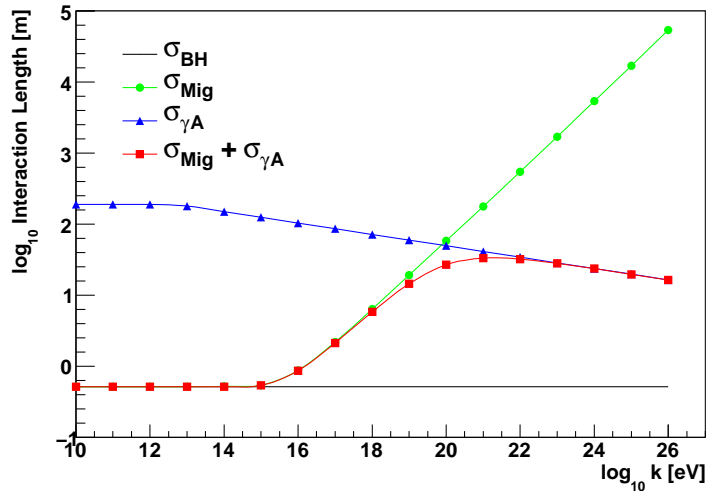}
\caption
{Interaction length of a photon in ice as a function of photon energy for the Bethe-Heitler (BH), Migdal (Mig), and photonuclear (${\gamma}A$) cross sections.}
\label{fig:ppintlen}
\end{figure}

\begin{figure} [tb]
\centering
\includegraphics[totalheight=0.3\textheight]{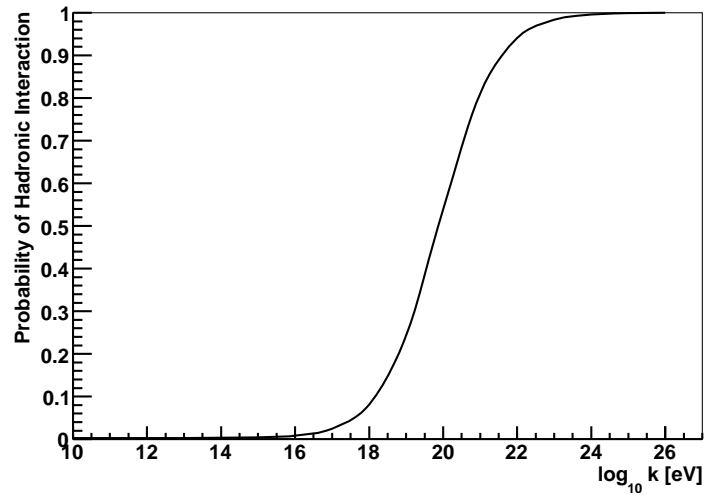}
\caption
{Probability of a photon interacting hadronically in ice as a function of photon energy.}
\label{fig:ppintprob}
\end{figure}

Figure \ref{fig:ppintlen} shows $l_I$ as a function of photon energy, and Figure \ref{fig:ppintprob} shows the probability for a photonuclear interaction. The length is a constant and the photonuclear interaction probability negligible
up to an energy of 4 PeV (about $15 E_{LPM}$)  when the LPM effect becomes significant and the interaction length rises.  $l_l$ reaches a maximum at a few $10^{21}$ eV, where the falling pair-production cross section meets the rising photonuclear cross section.  

We neglect radiative corrections to  pair production \cite{mork65}.  At sub-LPM energies, the radiative cross sections are typically a few percent of the non-radiative cross section.  Radiative processes have higher $q_{||}$ than their non-radiative counterparts, due both to the kinematics of the extra photon and the photon effective mass $\hbar\omega_p$.   A radiated photon with energy $r$ introduces a term $r\theta_r^2/2c$ to $q_{||}$, where $\theta_r$ is the angle between the initial and radiated photon directions, while the photon effective mass adds a term $(\hbar\omega_p)^2/2rc$; neither of these terms is a major correction to $q_{||}$, so radiative processes remain negligible even when LPM suppression is strong. 

\section{Bremsstrahlung cross sections}

Although Migdal's formulation made a number of improvements, including replacing the average scattering angle with a realistic distribution of scattering as well as considering quantum effects such as electron spin and photon polarization, it cannot easily include the suppression of bremsstrahlung when the produced photon interacts. In order to include all of these effects we equate Migdal's cross section with $S_{LPM+Di}*\sigma_{BH}$ and the bremsstrahlung cross section is given by the product of $(S_{All}/S_{LPM+Di})$ and Migdal's cross section (including dielectric suppression) shown in Eq. (\ref{eqn:migdie}).

For the semi-classical calculation, we sum the $q_{||}$ from different sources described in sections \ref{sect:two} and \ref{sect:three},  
\begin{equation}\label{eq:allqp}
q_{||}=\frac{m^2c^3k}{2E(E-k)}+\frac{k{\theta_{MS/2}}^2}{2c}+\frac{({\hbar\omega_p})^2}{2ck}+{\hbar}n({\sigma}_{PP}+{\sigma}_{\gamma A})
\end{equation}

\begin{figure} [tb]
\centering
\includegraphics[width=0.5\textwidth]{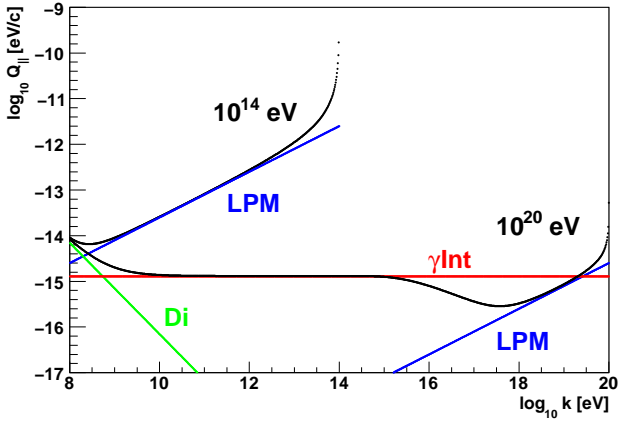}
\caption{The momentum transfer as a function of photon energy for $10^{14}$ and $10^{20}$ eV electrons in ice. The colored lines show the expected behavior for multiple scattering (LPM), photon interactions with the medium (Di), and photon interaction during the formation length ($\gamma$Int).  The different scaling regimes are discussed in Table \ref{tab:scaling}.  The minimum in $q_{||}$ for $k=10^{18}$ eV, $E=10^{20}$ eV occurs when the effects of multiple scattering begin to dominate over the effects of the photon interacting within the formation length.} 
\label{fig:qpar}
\end{figure}

Different terms in this equation dominate in different kinematic regimes; this leads to a rather complex behavior, with many different scaling laws. Figure \ref{fig:qpar} shows $q_{||}$ as a function of photon energy for two different electron energies.   Table \ref{tab:scaling} shows the five different scaling laws that are visible in the plot, along with the corresponding photon energy ranges.  
In each regime, the bremsstrahlung photon spectrum follows a different power, $d\sigma/dk\approx k^n$.  

For high enough electron energies, when the radiated photon is energetic enough that it interacts hadronically, a sixth law emerges.  Both the photon interaction cross section and the relevant term in $q_{||}$ rise as $k^{0.08}$.  In the regime where this term dominates $q_{||}$,  the bremsstrahlung cross section scales as $d\sigma/dk\propto1/k^{0.08}$.  This requires $k>10^{20}$ eV, and this scaling only applies for electron energies $E>10^{23}$ eV. 

\begin{table*}
\begin{tabular}{|l|c|c|c|c|}
\hline
Regime 	& $q_{||}$  		& Index $n$ for  			& $k$ range  	           &$k$ range  \\
		& Dependence 	& d$\sigma/dk\approx k^n$ 	&  ($E=10^{14}$ eV) 	  & ($E=10^{20}$ eV) \\
\hline
Bethe-Heitler 			& $k$ 		        & $-1$ & $> 1.4 \times 10^{13}$ 	           & $\approx 10^{20*}$\\
LPM 			& $k^{1/2}$	        & $-1/2$ & $2 \times 10^{8} - 1.4 \times 10^{13}$   & $> 3 \times 10^{17}$ \\
Dielectric (Di)		& $1/k$	& $1$	        & $< 2 \times 10^{8}$			   & $< 10^{8}$ \\
$\gamma$Int (pair no LPM)	& $k^{-1/2}$	        & $1/2$ & absent			           & $1.7 \times 10^{16} - 3 \times 10^{17}$ \\
$\gamma$Int (pair LPM)		&$k^0$	& $0$ 	        & absent			           & $5.5 \times 10^8 - 1.7 \times 10^{16}$ \\

\hline
\end{tabular}
\caption{The different photon-energy dependencies shown in Figure \ref{fig:qpar}.  ``Regime'' refers to the largest contribution to $q_{||}$ in Eq. (\ref{eq:allqp}).  The third column shows the power $n$  for photon spectrum in the bremsstrahlung cross section, $d\sigma/dk\approx k^n$. ``$\gamma$Int'' refers to the regime where $q_{||}$ is dominated by photon pair conversion.  It has
two subclasses, depending on if the photon energy is high enough that LPM suppression is important for the pair conversion. 
$^*$For $10^{20}$ eV electrons, the Bethe-Heitler regime holds as long as the photon carries away 99.9999\% of the electron energy, leaving the electron with less than $10^{14}$ eV.   Contributions from photonuclear interactions to $q_{||}$ start to become important at electron energies above $10^{23}$ eV.}
\label{tab:scaling}
\end{table*}

\begin{figure}[tb]
\begin{center}
\subfigure{
\includegraphics[width=1\columnwidth]{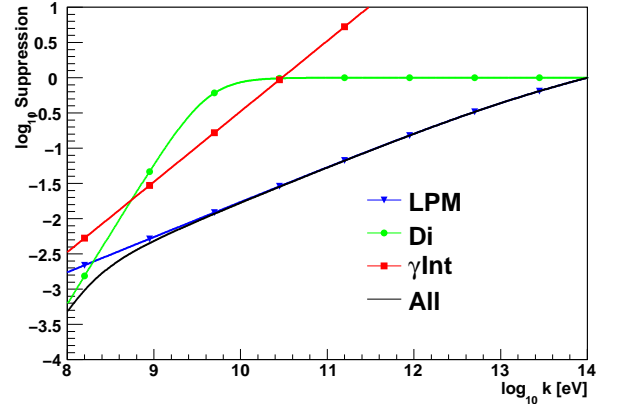}}
\subfigure{
\includegraphics[width=1\columnwidth]{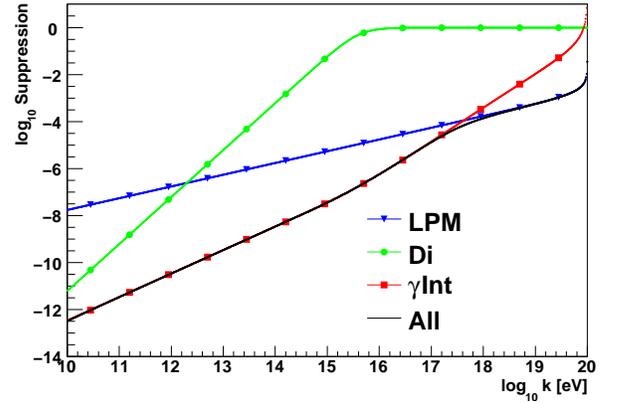}}
\caption{\label{fig:suppmsdiepp}
Suppression due to multiple scattering (LPM), the photon interaction with the medium (Di), and the photon interaction within the formation length (${\gamma}$Int)  for a 10$^{14}$ eV (top) and 10$^{20}$ eV (bottom) electron in ice.}
\end{center}
\end{figure}

Equation (\ref{eq:allqp}) retains the quadratic $l_f$ dependence from Section II; now the total suppression factor is 
\begin{eqnarray*}
S_{All}&=&\frac{-(1\!+\!A_{Di}\!+\!A_{{\gamma}Int})\!\pm\!\sqrt{\!(\!1\!+\!A_{Di}\!+\!A_{{\gamma}\!I\!n\!t})^2\!+\!4A_{L\!P\!M}}}{2A_{L\!P\!M}}\\
A_{L\!P\!M}&=&\frac{k{l_0}^2{E_s}^2}{4c{\hbar}X_0E(E-k)}\\
A_{Di}&=&\frac{{\hbar}{\omega_p}^2l_0}{2ck}\\
A_{{\gamma}\!I\!n\!t}&=&{l_0}n({\sigma}_{PP}+{\sigma}_{\gamma A})
\end{eqnarray*}
where $l_0=2{\hbar}E(E-k)/m^2c^3k$. Figure \ref{fig:suppmsdiepp} shows the suppression due to multiple scattering, the photon interacting with the medium, and photon interaction within the formation length as a function of photon energy for two different electron energies in ice. 

\section{Other Electron Interactions}

In the kinematic region where bremsstrahlung is strongly suppressed, two processes become important: direct pair production, $e^-N\rightarrow e^+e^-e^-N$, and electronuclear interactions, whereby the electron radiates a virtual photon which then interacts hadronically with a nearby nucleus.  The cross sections for both of these processes rise with increasing $E$.  

\subsection{Direct Pair Production}

The cross section for direct pair production, $eN\rightarrow e^+e^-eN$  was calculated long ago \cite{bhabha35,block54}.  These early calculations did not  include LPM suppression.    We use a recent calculation that does include LPM suppression  \cite{baier08}.   The cross section follows Eq. (5) of Baier and Katkov \cite{baier08} with modifications to use consistent values of $X_{0}$ and $E_{LPM}$ for ``L'' and ``$\varepsilon_{e}$'' and with the definition of ``$\nu_{1}$'' from \cite{baier05}.  LPM suppression is only considered for soft radiation (pair energy much less than incident electron energy), but we know of no better cross sections.  

\begin{figure}[tb]
\begin{center}
\includegraphics[width=0.5\textwidth]{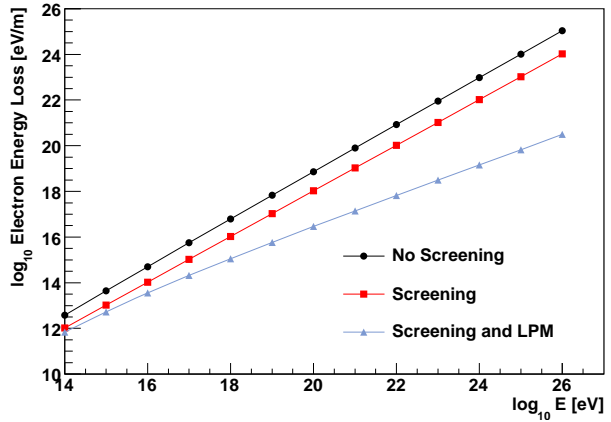}
\caption{The electron energy loss due to various predictions for direct pair production. The curve labeled ``No Screening'' is from Ref. \cite{block54}, the curve labeled ``Screening'' is from Ref.  \cite{bhabha35}, and the curve labeled ``Screening and LPM'' is from Ref. \cite{baier08}. See text for details.}
\label{fig:electronelosspp}
\end{center}
\end{figure}

Figure \ref{fig:electronelosspp} compares the electron energy loss for the Baier and Katkov cross section with two other cross sections which don't include LPM effects. The unscreened cross section is from Eq. (3) of Block {\it et al.} \cite{block54} and is based on a modification of the cross section given by Bhabha \cite{bhabha35}. The second cross section is from Eq. (46) of Bhabha \cite{bhabha35} and includes the effects of screening at high electron pair energies.  The Baier and Katkov curve agrees well with the screening calculation in the regime where LPM suppression is small.  At higher energies, the LPM effect reduces the energy loss; at $10^{20}$ eV, the energy loss is a factor of about 10 lower. 

\subsection{Electronuclear Interactions}

Electronuclear interactions occur when an electron emits a virtual photon, which then interacts hadronically with the target medium.  
The cross section for this is given by the Weizs\"acker-Williams method: 
\begin{equation}
\sigma_{eA}(E,k) = \int_{k_{min}}^E \frac{dN_\gamma}{dk} \sigma_{\gamma A}dk.
\label{eq:electronuclearflux}
\end{equation}
The choice of low energy cutoff, $k_{min}$, is not too important; we use $k_{min}=10^{-5}E$. This choice avoids the effects of nuclear or hadronic resonances. 
The photon flux $dN_\gamma/dk$ from an electron is from Eq. (71) of Ref. \cite{kossov02}:
\begin{equation}
\frac{dN_\gamma}{dk} = \frac{2\alpha}{\pi k} 
\bigg[ \frac{(2E-k)^2 + k^2}{2E^2} \ln(\gamma) - 1\bigg].
\label{eq:electronuclearflux1}
\end{equation}

We  use the same cross sections that we used to compute the photon interactions.   Although Eq. (\ref{eq:electronuclearflux1}) includes the effects of photon virtuality, the cross section does not account for this.   The other uncertainties in the calculation  (such as in $\sigma_{\gamma A}$) dwarf the inaccuracy associated with ignoring the photon virtuality. 

Equation (\ref{eq:electronuclearflux1}) is calculated without screening.  However, photonuclear interactions can occur out to impact parameters
\begin{equation}
b_{max} = \frac{\gamma\hbar c}{k}.
\end{equation}
At small $y$,  $b_{max}$ can be larger than the Thomas-Fermi screening radius, $a= 121 Z^{-1/3} \hbar/(m c) = 4.68\times 10^{-11} {\rm m}  \times Z^{-1/3}$ (defined in Ref. \cite{tsai74} just below Eq. (B24)).  Then, screening effects must be considered.  When $k< \gamma\hbar c/a$,  the logarithm in Eq. (\ref{eq:electronuclearflux1}) should be replaced with $\ln(184Z^{-1/3})$ to account for the screening of the electromagnetic fields at distances greater than $a$ from the electron.   There is a significant (factor of 2) discontinuity between
these two formulae at $k = \gamma\hbar c/a$; this occurs for $y=8\times10^{-3}$ for hydrogen and $y=1.7\times10^{-2}$ for oxygen.  After integration over $k$, the effect of screening is small, but it would affect a measurement of electron $dE/dx$.

\begin{figure}[tb]
\begin{center}
\includegraphics[width=1\columnwidth]{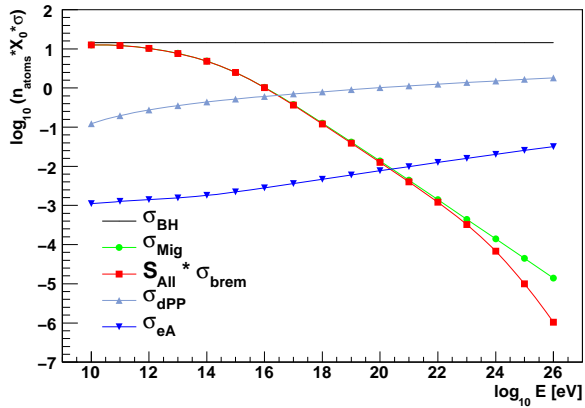}
\caption{The bremsstrahlung cross sections from Bethe-Heitler (BH), Migdal (Mig), and with all suppression effects ($S_{All} \times \sigma_{brem}$), as well as the cross sections for direct pair production (dPP), and electronuclear interactions (eA) as a function of electron energy (integrated over photon energies above $10^{-5}$E) for an electron traveling through ice.}
\label{fig:electronsigma}
\end{center}
\end{figure}

Figure \ref{fig:electronsigma} compares the cross sections (integrated over $k$) for bremsstrahlung with and without suppression effects, direct pair production, and electronuclear interactions in ice.  At energies above $10^{16}$ eV, direct pair production has the largest cross-section. 

Although we do not consider it here, coherent $\rho^0$ electroproduction is also a possibility; this is analagous to coherent $\rho^0$ photoproduction \cite{couderc09}.  

\section{Electron Interaction Length}

\begin{figure}[tb]
\begin{center}
\includegraphics[width=0.5\textwidth]{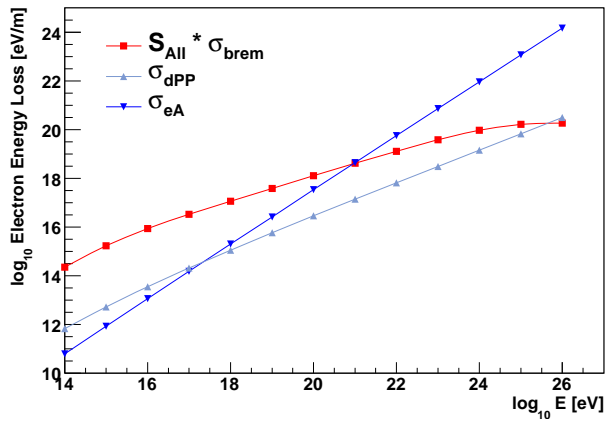}
\caption{The electron energy loss due to suppressed bremsstrahlung, electronuclear interactions, and direct pair production as a function of electron energy.}
\label{fig:electroneloss}
\end{center}
\end{figure}

Figure \ref{fig:electroneloss} shows the electron energy loss for suppressed bremsstrahlung, direct pair production, and electronuclear interactions. The total electron interaction length depends on the total energy loss from these processes.  We quantify it with  
\begin{equation}
X_{TOT} = \frac{E}{\int_{k_{min}}^E  k n d\sigma/dk},
\label{eqn:eint}
\end{equation}
keeping $k_{min}= 10^{-5}$E.    $X_{TOT}$ is dominated by the region of large k, $k>0.1 E$, so it is somewhat less sensitive to some of the effects discussed above.  These losses are similar to those found in Fig. 1 of Ref. \cite{ralston02}.

At electron energies above $10^{21}$ eV, electrons lose energy mostly by electronuclear interactions. This is similar to what happened with the photon interactions at similar energies. At higher energies, further LPM suppression of bremsstrahlung (or direct pair production) becomes unimportant in shower development and the character of the interactions changes. Instead of a single, very hard bremsstrahlung (hard because LPM suppression eliminates the soft radiation), electrons lose energy in a series of lower-energy interactions.  The increase in the number of interactions can be estimated from the cross section ratios in Figure \ref{fig:electronsigma}. At an energy of $10^{21}$ eV, where the energy loss is similar, the direct pair production cross section is about two orders of magnitude higher than the bremsstrahlung cross section, while the electronuclear cross section is about three times larger.    Most of the produced particles are $e^+e^-$ pairs, but the bulk of the energy is lost hadronically.    Between the higher cross sections and the higher hadronic multiplicity, the event-to-event fluctuations should be much smaller than  in purely electromagnetic (only bremsstrahlung and pair production) calculations \cite{klein04,konishi91}. In short, ultra-high energy electrons electrons radiate much like muons \cite{mmc}.

Hadronic showers are not affected by LPM suppression \cite{zas98} and the interaction energy is deposited in a relatively compact volume. Because the transverse momentum scale in hadronic interactions is larger than in electromagnetic showers, the transverse size of the shower may be larger, shifting radio emission to lower frequencies. 

Muons produced in the interactions can travel long distances \cite{klein04}. In the Rogers \& Strikman dipole model, charm production is significantly enhanced at high energies; at a photon energy of $10^{21}$ eV, the leading charm fraction is 25-30\%, depending on target composition \cite{rogers06}.  Even though most ultra-high energy charmed hadrons will interact and lose energy before they decay, prompt muon production will still be significant. 

\begin{figure}[tb]
\begin{center}
\includegraphics[width=0.5\textwidth]{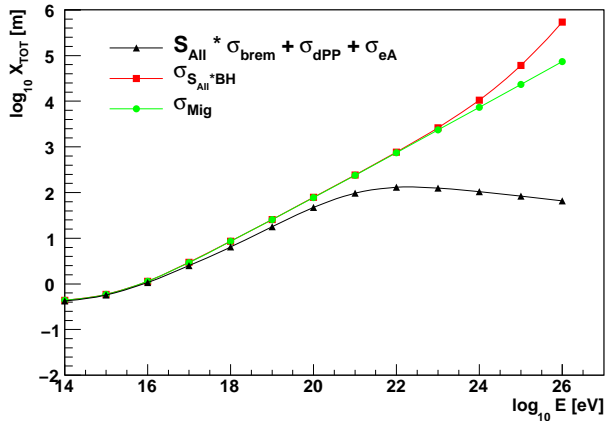}
\caption{The electron interaction length (Eq. (\ref{eqn:eint})) due to bremsstrahlung as a function of electron energy in ice.}
\label{fig:electronlosslength}
\end{center}
\end{figure}

The total electron interaction length is shown in Figure  \ref{fig:electronlosslength}.  When all of the processes discussed here are included the electron interaction length reaches a maximum at an energy around $10^{21}$ eV and decreases slowly as electron energy increases further.  

The largest uncertainty in the energy loss is from the uncertainties in the hadronic cross sections; these energies require significant extrapolation from existing data, and different methods give different results.  For example, a Glauber calculation \cite{KN99} predicts a photonuclear cross section a factor of 2 smaller than the Engel parameterization \cite{couderc09}.  A factor of 2 change in cross section will shift the cross-over energy  between bremsstrahlung and electronuclear interactions by 40\%.

\section{Electromagnetic Shower Development}

The length of electromagnetic showers produced in $\nu_e$ interactions has often been calculated
 with a toy model for shower development \cite{ros41}.  The created  electron initiates an electromagnetic shower which evolves via bremsstrahlung and pair production, with each succeeding shower generation containing twice as many particles as the last, each with half the energy. Shower development continues until the average particle energy reaches the critical energy, $E_c$, the value where Compton scattering dominates. In ice, $E_c=79$ MeV. The interaction length is  the average of the electron and photon interaction lengths and each generation occupies a length $\ln{(2)} X_0$ because $X_0$ is the distance for a particle to lose all but $1/e$ of its energy. 

When direct pair production and electronuclear interactions are included, this model becomes problematic because the electron can interact hadronically, spreading the lost energy among many particles. To include this, we calculate the shower length with the integral
\begin{equation}
L_{EM}=\int_{E_{c}}^E \frac{X_{TOT}(E)}{E} dE
\label{eqn:showerlen}
\end{equation}
where $X_{TOT}$ includes all of the energy loss processes.  This is essentially the integral of the inverse of the electron energy loss. When only bremsstrahlung and pair production are considered (i.e. where $X_{TOT} = X_{0}$), it gives a similar result to the splitting method discussed above: $X_{0} \ln(E/E_{c})$.

Figure \ref{fig:showerlength} shows the shower length calculated for bremstrahlung and pair production cross sections without supression (Beithe-Heitler: Eqs. (\ref{eq:BH2}) and (\ref{eq:BH3})), with LPM and dielectric suppression (Migdal: Eqs. (\ref{eq:LPM}) and (\ref{eqn:migpair})), and with all of the photon and electron suppression effects, all calculated with Eq. {\ref{eqn:showerlen}.  These suppressed values also includes the effects of the electronuclear, direct pair production, and photonuclear cross sections.  

For comparison, Fig. \ref{fig:showerlength} also shows the hadronic shower length, using the approach of Refs. \cite{klein04,ros41}.  The hadronic interaction length shown in Figure \ref{fig:showerlength} is calculated using a method similar to the one used for electromagnetic showers.  Every hadronic generation has twice as many particles, each with half the energy, until the critical energy is reached. These generations occupy one hadronic interaction length,  90.7 cm  in ice.  As is noted in Ref. \cite{klein04}, this approach may over-estimate the shower length by 30-70\%. At low energies, hadronic showers are longer than electromagnetic, largely because of the longer hadronic interaction length. The hadronic shower length rises slowly throughout the entire energy range and, for energies above $10^{19}$ eV, hadronic showers are more compact than electromagnetic. 

At energies below $10^{16}$ eV, the three electromagnetic curves are almost identical, and the shower length increases  roughly logarithmically with $\nu_e$ energy, $E_\nu$.  Above $10^{16}$ eV, the shower lengths with LPM suppression increase more rapidly, as $\sqrt{E_\nu}$, reaching a length of about 200 m at $10^{20}$ eV.   At still higher energies, the purely electromagnetic (bremsstrahlung and pair production only) shower length continues to rise as $\sqrt{E_\nu}$. In contrast, when photonuclear and electronuclear interactions are included, the increase in shower length levels off, returning to near the $\log{(E_\nu)}$ behavior.   At these high energies, there will also be a long-range (10s of kilometers) tail of energy deposition from prompt muons. 

At $10^{24}$ eV, these electron shower lengths are about seventy times larger than without LPM suppression, and about an order of magitude shorter than the purely electromagnetic shower with LPM suppression.  Still, at a total shower length of $\sim$1 km, they are only about thirty times larger than their corresponding hadronic shower lengths so it should be possible to make use of the radio emission from these showers.  Since the electron shower contains the bulk of the $\nu_e$ energy, inclusion of this component could lead to a significant decrease in experimental energy threshold. 

This shower length comparison only applies for electron initiated showers, including those from $\nu_e$.   It should also be appropriate for photons with energies below about $10^{20}$ eV.     Higher energy primary photons interact hadronically, depositing all of their energy in a relatively compact shower, and so the average shower length for photons may be found by adding  the photon interaction length in Fig. \ref{fig:ppintlen} and the hadronic shower length in Fig. \ref{fig:showerlength}.

\begin{figure}[tb]
\begin{center}
\includegraphics[width=0.5\textwidth]{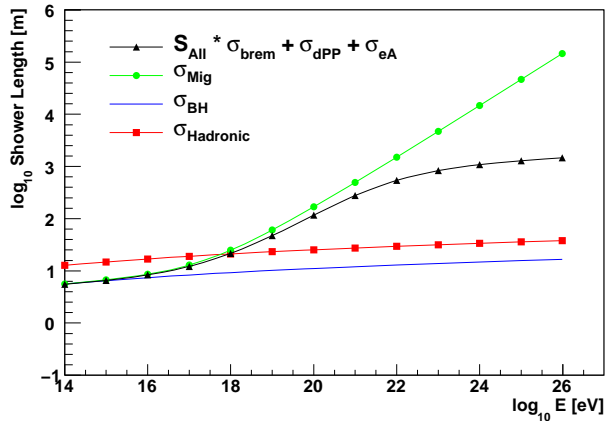}
\caption{The length of an electromagnetic and hadronic shower as a function of electron energy in ice.}  
\label{fig:showerlength}
\end{center}
\end{figure}

\section{Conclusions}

We have performed detailed  calculations of bremsstrahlung and pair production at high energies, where the LPM effect is strong.  We consider the mutual suppression of pair production and bremsstrahlung.  Pair production of the nascent photons reduces the bremsstrahlung cross section in ice at energies above 1 PeV, but this does not alter the overall shower development.

At higher energies, above $10^{20}$ eV in ice, photonuclear and electronuclear interactions become dominant contributors to photon interactions and electron energy loss.   In this energy range, these two reactions control shower development, further reductions of the bremsstrahlung and pair production cross sections are unimportant, and the total cross section grows slowly with energy.   

The changing interaction length drives the total shower length.  At energies below $10^{16}$ eV, the total shower length rises slowly, roughly logarithmically, with energy.  Above this threshold, the LPM effect becomes important and the shower length rises roughly as $\sqrt{E}$.  Above $10^{20}$ eV, photonuclear and electronuclear interactions become important, and the shower length increases only slowly with increasing neutrino energy.  At the highest energies, electron-initiated showers are only about a factor of thirty longer than hadronic showers, and so should still emit useful fluxes of radio waves.  Since $\nu_e$ transfer the bulk of their energy to the produced electron, consideration of this radiation should lead to improved sensitivity for future cosmic neutrino searches. 

At energies above $10^{20}$ eV, the behavior of electron and photon induced showers diverges.  Photons interact hadronically, producing a compact shower containing all of the initial state energy, while electrons lose energy gradually, depositing energy over longer distances.  In both cases the shower develops hadronically, so it is wider and contains more muons than a purely electromagnetic shower. 

We thank Brian Mercurio for helpful discussions. This work was funded by the U.S. National Science Foundation under grant 0653266 and the U.S. Department of Energy under contract number DE-AC-76SF00098.


\begin{thebibliography}{99}

\def\etal{{\it et al.}}

\bibitem{IceCube}M. Ackermann \etal, Astrophys. J. {\bf 675}, 1014 (2008). 

\bibitem{RICE}I. Kravchenko \etal, Phys. Rev. {\bf D73}, 082002 (2006).

\bibitem{ARIANNA}L. Gerhardt {\it et al.}, preprint arXiv:1005.5193; S. Barwick, Nucl. Instrum. \& Meth. {\bf A602}, 279 (2009).

\bibitem{ARA}P. Chen and K. Hoffman, preprint arXiv:0902.3288.

\bibitem{ANITA}P. W. Gorham \etal, preprint arXiv:1003.2961; P. W. Gorham \etal, Phys. Rev. Lett. {\bf 103}, 051103 (2009).

\bibitem{Acoustic}J. Vandenbroucke, preprint astro-ph/0611503.

\bibitem{Parkes}T. Hankins \etal, Monthly Notices Royal Astronomical Society {\bf 283}, 1027 (1996).

\bibitem{GLUE}P. Gorham \etal, Phys. Rev. Lett. {\bf 93}, 041101 (2004).

\bibitem{numoon}S. Buitink \etal, preprint arXiv:1004.0274.

\bibitem{Lunaska}C. W. James \etal, Phys. Rev. {\bf D81}, 042003 (2010).

\bibitem{RESUN}T R. Jaeger, R. L. Mutel and K. G. Gayley, preprint arXiv:0910.5949.

\bibitem{Lofar}J. Horandel \etal, Nucl. Phys. B (Proc. Suppl.) {\bf 196}, 289 (2009).

\bibitem{klein04}S. R. Klein, preprint astro-ph/0412546.

\bibitem{konishi91}E. Konishi \etal, J. Phys. G. {\bf 17}, 719 (1991); A. Misaki, Nucl. Phys. B Proc. Suppl. {\bf 33A,b}, 192 (1993); S. Klein, astro-ph/9712198; J Bolmont {\it et al.}, in arXiv:0711.0353.

\bibitem{ralston02}J. P. Ralston, S. Razzaque and P. Jain, preprint astro-ph/0209455.

\bibitem{bet43}H. A. Bethe and W.Heitler,   Proc. R. Soc. London, Ser. , {\bf 146}, 83 (1943).

\bibitem{tsai74}Y. S. Tsai, Rev. Mod. Phys. {\bf 46}, 815 (1974). 

\bibitem{land53}L.~Landau and I.~Pomeranchuk.Dokl. Akad. Nauk SSSR, {\bf 92}, 735 (1953).
\newblock English translation: L.~Landau, {\em The Collected Papers of L.D.~Landau} (Pergamon, New York, 1965), p. 589.

\bibitem{anthony9598}P. Anthony \etal, Phys. Rev. Lett. {\bf 75}, 1949 (1995); P. Anthony \etal, Phys. Rev. {\bf D56}, 1373 (1997). 

\bibitem{hansen04}H. D. Hansen \etal, Phys. Rev. {\bf D69}, 032001 (2004). 

\bibitem{spen99}S.~R. Klein. Rev. Mod. Phys. {\bf  71}, 1501 (1999).

\bibitem{ter53}M.L. Ter-Mikaelian, Zh. Eksp. Teor. Fiz. {\bf  25}, 289  (1953); M.L. Ter-Mikaelian, Zh. Eksp. Teor. Fiz. {\bf 25}, 296  (1953).

\bibitem{feinberg56}E. I. Feinberg and I. Pomeranchuk, Nuovo Cim. Suppl. A1, Serie X{\bf III}, 652 (1956).

\bibitem{mig56}A. B. Migdal, Phys. Rev. {\bf 103}, 1811 (1956). 

\bibitem{stan82}T.~Stanev \etal. Phys. Rev. {\bf D, 25}, 1291 (1982).

\bibitem{baier00}V. N. Baier and V. M. Katkov, Phys. Rev. {\bf D62}, 036008 (2000).

\bibitem{zakharov99}B. G. Zakharov, JETP Lett. {\bf 64}, 781 (1996).

\bibitem{klein06}S.~Klein,  Radiation Physics and Chemistry {\bf 75}, 696 (2006).

\bibitem{baier03}V. N. Baier and V. N. Katkov, Phys. Rept. {\bf 409}, 261 (2005). 

\bibitem{gal64}V. M. Galitsky and I.I. Gurevich, Nuovo Cimento, {\bf 32}, 396 (1964).

\bibitem{engel97}R. Engel, J. Ranft and S. Roesler, Phys. Rev. {\bf D55}, 6957 (1997).

\bibitem{couderc09}E. Couderc and S. R. Klein, Phys. Rev. Lett. {\bf 103}, 062504 (2009).  See also the comment on this paper
by T. C. Rogers and M. Strikman, Phys. Rev. Lett. {\bf 103}, 259201 (2009) and the response, 
E. Couderc and S. Klein, Phys. Rev. Lett. {\bf 103}, 259202 (2009).

\bibitem{rogers06}T. Rogers and M. Strikman, J. Phys. G {\bf 32}, 2041 (2006).

\bibitem{KN99}S. Klein and J. Nystrand, Phys. Rev. {\bf C60}, 014903 (1999).

\bibitem{mork65}K. Mork and H. Olsen, Phys. Rev. {\bf 140}, B1661 (1965). 

\bibitem{bhabha35}H. J. Bhabha, Proc. Royal Soc. of London A {\bf 152}, 559 (1935).

\bibitem{block54}M. M. Block, D. T. King and W. W. Wada, Phys. Rev. {\bf 96}, 1627 (1954).

\bibitem{baier08}V. Baier and V. M. Katkov, JETP Lett. {\bf 88}, 80 (2008). 

\bibitem{baier05}V. Baier and V. M. Katkov, Phys. Repts. {\bf 409}, 261 (2005).

\bibitem{kossov02}M. V. Kossov, Eur. Phys. J. {\bf A14}, 377 (2002).

\bibitem{Jackson}J. D. Jackson, {\it Classical Electrodynamics}, 3rd edition, Wiley, 1998.

\bibitem{mmc}D. Chirkin and W. Rhode, preprint hep-ph/0407078; S. Polityko {\it et al.}, J. Phys. G: Nucl. Part. Phys. {\bf 28}, 427 (2002).

\bibitem{zas98}J. Alvarez-Muniz and E. Zas, Phys. Lett. {\bf B434}, 396 (1998).

\bibitem{ros41}B. Rossi and K. Greisen, Rev. Mod. Phys. {\bf 13}, 240 (1941).

\end{thebibliography}
\end{document}